\documentclass[fleqn,twocolumn]{revtex4}
\usepackage[utf8]{inputenc}
\usepackage[T1]{fontenc}

\usepackage{graphicx,graphics,color,epsfig}
\usepackage{amsmath}
\usepackage{amssymb}
\usepackage{times}

\begin{document}

\title{
Machine learning as an improved estimator for
magnetization curve and spin gap
}
\author{Tota Nakamura}
 \affiliation{%
Faculty of Engineering, Shibaura Institute of Technology, %
307 Fukasaku, Minuma, Saitama 337-8570, Japan}

\date{\today}

\begin{abstract}
The magnetization process is a very important probe to study 
magnetic materials, particularly in search of spin-liquid states
in quantum spin systems.
Regrettably, however,
progress of the theoretical analysis has been unsatisfactory, mostly
because it is hard to obtain sufficient numerical data to support the theory.
Here we propose a machine-learning algorithm 
that produces the magnetization curve and the spin gap well
out of poor numerical data.
The plateau magnetization, its critical field and the critical exponent
are estimated accurately.
One of the hyperparameters identifies by its score whether
the spin gap in the thermodynamic limit is zero or finite.
After checking the validity for exactly solvable one-dimensional models
we apply our algorithm to the kagome antiferromagnet.
The magnetization curve that we
obtain from the exact-diagonalization data with 36 spins 
is consistent with the DMRG results with 132 spins.
We estimate the spin gap in the thermodynamic limit at a very small 
but finite value. 
\end{abstract}

\maketitle

\section{Introduction}
Past studies have paid much attention to obtaining accurate
raw data; 
however, in our view,
devising a more accurate data-analysis method warrants more attention.
We still encounter an analysis that 
fits a straight line to data to reach a conclusion, despite that
the fitting may depend on a way how the data were plotted.
Recently,
the situation has been drastically changed 
by the introduction of machine learning~\cite{bishop,rinko,goodfellow}
to data analyses.
The neural-network algorithm has been one of the most successful method 
in various fields of science and technology;
we can list its many examples% 
~\cite{melko,atanaka,carleo,broecker,huber,ohtsuki}
applied to physical systems.
An advantage of the neural-network algorithm is the scalability;
the computational cost is of linear order of the data size,
thanks to which we can deal with big data easily.
The algorithm is generally robust because of the statistical 
stability of big data size.
Another type of machine-learning algorithm becomes available for 
the cases that the data size does not exceed one thousand, namely
the Gaussian-kernel method coupled
with the Bayesian inference~\cite{toussaint,harada}.
We employ it to study the magnetization process 
of quantum spin systems in this paper.

The magnetization process 
enables us to access the ground state of quantum spin systems
both experimentally~\cite{kageyama,tanaka2}
and 
theoretically~\cite{nishimori,chubukov,oshikawa}.
We can identify the type of the magnetic phase
and estimate the spin gap out of the magnetization curve.
We can fix physical constants
by comparing numerical results for the curve with experimental results.
Estimation of the critical exponent is also important to 
classify the universality class of the critical phenomenon.
In such studies on quantum spin systems,
a highlight is the search for the
spin-liquid state~\cite{balents,savary,anderson}, in which
magnetic spins remain disordered down to the zero temperature but
are strongly correlated with each other in some manner.
It exhibits a variety of exotic phenomena, including 
the Haldane state~\cite{haldane} in the $S=1$ antiferromagnetic spin chain and
the spin-liquid state in the Kitaev model~\cite{kitaev}.
In such phenomena, magnetic frustration 
in the spin-spin interactions plays an important role.
Thus, the ground state of frustrated quantum spin systems
has been attracting much interest.

The kagome antiferromagnet is a well-known prototype of the frustrated
quantum spin systems.
It has been investigated intensively for more than two decades.
However,
the existence of a finite
spin gap~\cite{waldtmann,jiang,sindzingre,nakano11b,yan,lauchli,%
depenbrock,iqbal,expkagome,heprx,ran,liao}
and
the location of the
magnetization plateaux in the magnetization process~%
\cite{hidaMH,schulen,honecker,cabra2,nakano10,sakai11,%
nishimotokagome,capponi,nakano14,picot,richter,nakano18,plat}
still remain under debate. 
This unsettling situation is partly due to the
lack of sufficient numerical verification of theoretical studies.
There is a serious technical problem in the
numerical study on frustrated quantum spin systems,
namely the notorious negative-sign problem 
in the quantum Monte Carlo simulation~\cite{qmcbook}.
Data sampling deteriorates exponentially with the increase of
the system size and with the decrease of 
the temperature~\cite{nakahata1,nakahata2,nakamiya}.
We cannot access the ground state of a large system
because of this problem.
An alternative method is the exact diagonalization (ED) but
its application is also restricted to small systems.
We have not found a promising numerical method to treat frustrated
systems up to the present.
Under these circumstances,
the development of a data-analysis method is particularly
important in order to extract a meaningful physical conclusion
out of poor numerical data.

The Gaussian-kernel method is a machine-learning algorithm
which enables analytically differentiable data regression.
Harada~\cite{harada} introduced 
the Bayesian inference coupled with the
Gaussian-kernel method for parameter estimation 
in the finite-size scaling analysis of critical phenomena.
It automatically evaluates the critical point and critical 
exponents without assuming any form of the scaling function.
Nakamura~\cite{totabayes} employed this method to find
continuous and analytically differentiable model functions of
physical quantities directly from raw simulation data.
For example, 
we obtained the internal energy as a continuous function of the temperature
as $E(T)$ out of energy data collected at discrete values of the temperature.
We then found the specific heat $C(T)$
as a continuous function by algebraic temperature 
differentiation of $E(T)$.
We also determined
the critical temperature as an exclusion point 
of $E(T)$ that cannot be modeled by a smooth function.
Its accuracy was within five digits 
of the exact value for the two-dimensional Ising model~\cite{totabayes}.
We further estimated the critical exponent by the logarithmic
differentiation of a model function of a physical quantity.
A similar approach~\cite{huber} 
was taken to estimate the critical temperature
as a point at which the neural network confuses the classification problem.

Empowered by this Bayesian inference coupled with the Gaussian-Kernel 
(B-GK) method, 
we acquire a useful tool to obtain a continuous and accurate estimate
of the magnetization curve 
in a style completely different from the conventional ones.
The magnetization plateau $M_{\rm p}$, 
its critical field $H_{\rm p}$, and its critical exponent $\delta$
will be estimated very accurately in a similar
manner to how we obtained the critical temperature and the critical exponent 
for the Ising model~\cite{totabayes}.
The method also gives an alternative estimate of the spin gap 
by the critical field of its zero-magnetization plateau.
Since its size dependence is different from 
the conventional spin-gap definition,
we can perform a combined size-extrapolation analysis 
using both series of data, which further
improves the accuracy of the estimate.
We also find that one hyperparameter in this analysis identifies
the ground state being gapless or gapful by its score.

For a check, we applied this method to
one-dimensional models, obtaining the results 
that agreed with the exact solutions.
Then, we used it to resolve the unsettled issues in the kagome antiferromagnet.
We estimated the spin gap in the thermodynamic limit at a
small but finite value.
The critical exponent of each magnetic plateau 
suggested that the excitations above the states 
on the finite-magnetization plateaux
as well as the excitations above the ground state
are different from those below the plateau states.

\section{Results}

\subsection{Validity check in one-dimensional models.}

We first tested the method for the magnetization curve in
the $S=1/2$ antiferromagnetic~(AF) Heisenberg
modulated spin chain~\cite{huincome},
\begin{equation}
{\cal H}=\sum_{i=1}^{N}  
(1-\lambda \cos[2\pi \alpha i + \phi])
{
\boldmath{S}_{i}\cdot \boldmath{S}_{i+1},
}
\label{eq:income}
\end  {equation}
where $N$ denotes the spin number, 
$\lambda$ is a dimerization parameter,
$\alpha$ is a modulation parameter,
and $\phi$ is a phase factor.
The periodic boundary condition is imposed.
We study the existence of an incommensurate or commensurate
ground state depending on the value of $\alpha$.
According to a criterion proposed by Oshikawa {\it et al.}~\cite{oshikawa},
we expect to observe a plateau in the magnetization curve at 
$M/N=S-\alpha$,
where $M$ denotes the total magnetization of the system and
$\alpha$ was originally set to a rational number.
Hu {\it et al.}~\cite{huincome} investigated this model with an irrational 
number of $\alpha$ 
in order to study the incommensurate ground state.
They obtained the plateau magnetization at $S-\alpha$
within the accuracy of $1/200$ when $\lambda=0.8$
by the density-matrix renormalization group (DMRG)
method applied up to $N=200$.
We here test our data-analysis method to estimate the plateau
magnetization using poor data obtained for
a system of much smaller size and with much smaller $\lambda$.

\subsubsection{The magnetization curve}

\begin{figure}
\hfil
  \resizebox{0.45\textwidth}{!}{\includegraphics{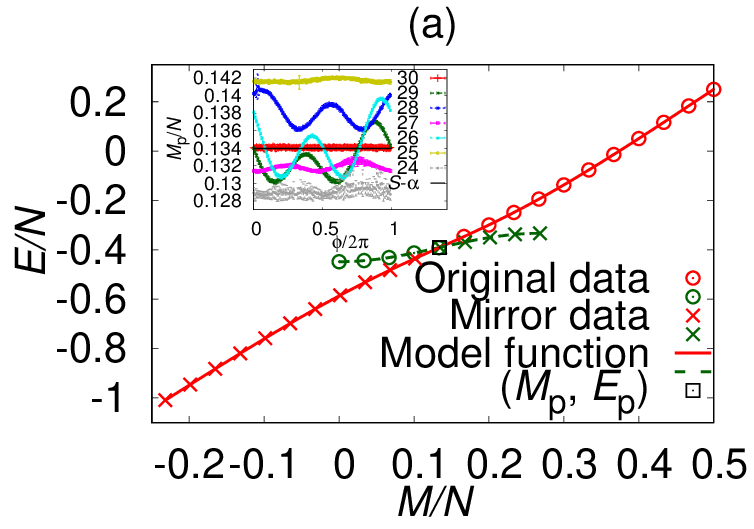}}
  \resizebox{0.45\textwidth}{!}{\includegraphics{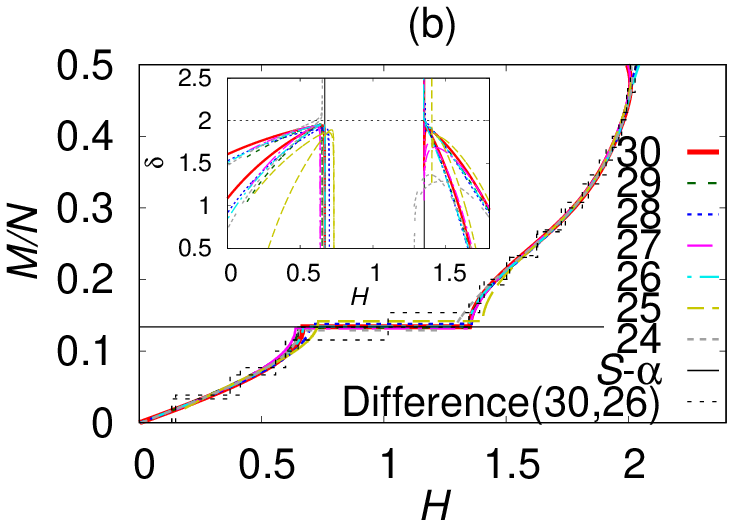}}
  \caption{
B-GK results for the modulated AF Heisenberg spin chain.
(a)
The ground-state energy $E(M)$ of a system 
$N=30$, $\alpha=(\sqrt{3}-1)/2$, $\lambda=0.2$, and $\phi=0$.
The original ED data, the estimated plateau point $(M_{\rm p}, E_{\rm p})$,
and mirror data are plotted with points.
Two model functions for the data above~(red) and below~(green)
the plateau point are plotted with lines.
The asymmetric parameters were estimated at 
$-0.84(9)$ and $-0.9994(2)$ for the upper and the lower 
side of the plateau point, respectively.
The inset shows the dependence of the estimated
plateau magnetization on the phase factor and the system size.
(b)
The magnetization curve for each system size ranging from $N=24$ to $30$
compared with the one obtained
by the difference for $N=26$ and $30$.
Our estimation of $M_{\rm p}/N= 0.1341(3)$ for $N=30$ is
consistent with $S-\alpha=0.1339745\cdots$.
The inset shows model functions of the exponent $\delta$ 
obtained by both expressions in equation~(\ref{eq:delta}).
We took the average over 100 phase factors and put an error of $M_{\rm p}$
by the standard deviation.
}
\label{fig:income}
\end{figure}

We calculated the ground-state energy for each value of the magnetization as
$E(M)$ by the ED method on a system of 30 spins; 
see Fig.~\ref{fig:income}(a).
Let
$E_{\rm p}$ the energy value at the plateau magnetization $M_{\rm p}$.
We fixed the plateau point $(M_{\rm p}, E_{\rm p})$
by the Bayesian inference as an exclusion point
at which the Gaussian-kernel method fails to model data by a smooth function.
If the magnetic plateau exists, the $E(M)$ data exhibit a jump of its
slope at $(M_{\rm p}, E_{\rm p})$ since $H=\partial E(M)/\partial M$.
We cannot model such data by a smooth function because the derivative
is not continuous.
In this inference, we introduced additional mirror data 
(depicted by crosses in Fig.~\ref{fig:income}(a))
corresponding to the original data 
(depicted by circles in Fig.~\ref{fig:income}(a))
for both upper and lower sides of the plateau point.
Locations of the mirror data are controlled by an 
additional hyperparameter
$A_{\rm s}$, which we hereafter call the asymmetric parameter.
The mirror data are located symmetrically or antisymmetrically 
when $A_{\rm s}=1$ or $-1$, respectively.
This asymmetric parameter is also fixed by the Bayesian inference.
We will below use the estimate of the asymmetric parameter as an
indicator of the existence of a finite spin gap.

By using estimated values of $(M_{\rm p}, E_{\rm p})$ and hyperparameters,
the B-GK method gives
a model function for $E(M)$ smoothly connecting the original data 
and the mirror data as shown in Fig.~\ref{fig:income}(a).
The present algorithm evaluates the value of $M_{\rm p}$ within 
five-digit accuracy.
This lets us clearly observe the phase-factor dependence of $M_{\rm p}$
due to the finite-size effects;
see an inset of Fig.~\ref{fig:income}(a).
This dependence is regarded as an outcome of the inconsistency between the
incommensurate modulation and the periodic boundary condition.
The magnetic interactions are not continuous at the periodic boundary edge 
if $\alpha N$ is not an integer.
When $N=30$, this inconsistency is relaxed because 
$\alpha N$ ($\simeq 10.98\cdots$) is close to an integer.
Therefore, the results did not show the $\phi$ dependence.

We draw the
magnetization curve by analytic differentiation:
$H(M)=\partial E(M)/\partial M$.
Note again that this is possible because we obtain $E(M)$ as an
analytic function except for at $M_{\rm p}$.
The results for various system sizes are
shown in Fig.~\ref{fig:income}(b).
Each B-GK result exhibits a continuous magnetization curve,
whose plateau magnetization is consistent with the expected value 
according to Oshikawa {\it et al.}'s 
criterion within four-digit accuracy when $N=30$.
It is a remarkable consistency considering the size of the system.
A value of $M_{\rm p}$ for each system size oscillates around and approaches 
$S-\alpha$ as the system size increases.
Values of the critical field $H_{\rm p}$ for both plateau edges 
also exhibit similar size dependences.
In addition to $M_{\rm p}$,
the point at $M/N=0.5$ is another data edge.
Since there is no numerical data above this point, contributions 
of the Gaussian kernel function from the upper side of the data are missing
in the vicinity of $M/N=0.5$. 
Then, the quality of the model function deteriorates and the finite-size
dependence appears.

We compared our results with the magnetization curve
estimated by the conventional method, that is, 
by taking the difference of the same data as
$H(M)=E(M)-E(M-1)$, which
exhibits a stepwise behavior.
Obviously,
we cannot estimate the incommensurate magnetization plateau accurately
from this plot.

A model function for the critical exponent $\delta$ 
defined by $ (H-H_{\rm p})\sim (M-M_{\rm p})^{\delta}$
is plotted in an inset of Fig.~\ref{fig:income}(b).
Here, we plotted two model functions given by two equivalent expressions 
for $\delta$ in equation~(\ref{eq:delta}).
Both model functions approach
two at the critical field, which is
consistent to the critical exponent in gapful systems~\cite{takahashi}.
Numerical instability occurred in the vicinity of $M_{\rm p}$.
We need to discard these data where $|M-M_{\rm p}|$ is smaller than 
several times the standard deviation of $M_{\rm p}$, 
which is estimated by
40 best results out of 800 trials of the Bayesian inference.

\subsubsection{The spin gap}

We also checked the present method in two exactly solvable models
particularly focusing on the existence of the spin gap:
\begin{equation}
{\cal H}=\sum_{i=1}^{N}  
(1+(-1)^i\lambda)
{
\boldmath{S}_{i}\cdot \boldmath{S}_{i+1}.
}
\label{eq:1d}
\end  {equation}
One is the XY-spin version of the Hamiltonian (\ref{eq:1d}), 
namely a dimerized AF XY chain.
We applied the magnetic field in the $z$ direction.
The other one is the uniform AF Heisenberg spin chain with $\lambda=0$.
The exact expression for the magnetization curve is
known~\cite{kontorovich,perk,okamoto,grif,takahashi} in both cases.
The magnitude of the spin gap in the thermodynamic limit 
is $\lambda$ in the XY model.
Therefore, we can tune the system gapful or gapless
by varying the dimerization parameter.
The ground state of the uniform AF Heisenberg spin chain is known 
to be gapless.

\begin{figure}[h]
\hfil
  \resizebox{0.40\textwidth}{!}{\includegraphics{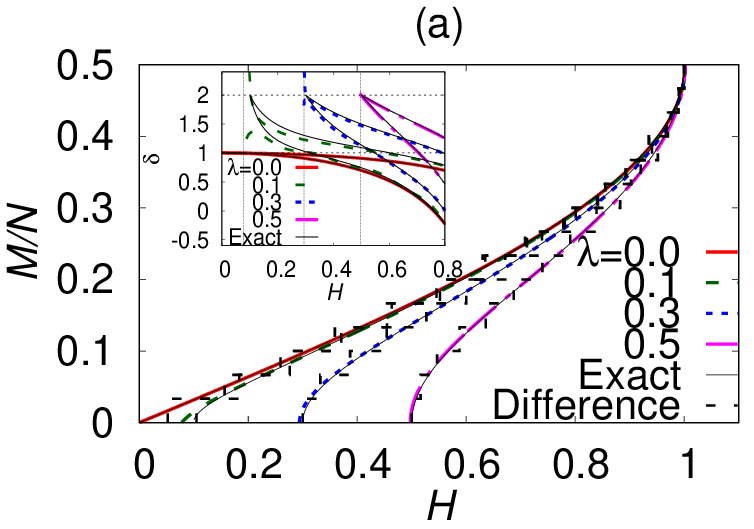}}
  \resizebox{0.40\textwidth}{!}{\includegraphics{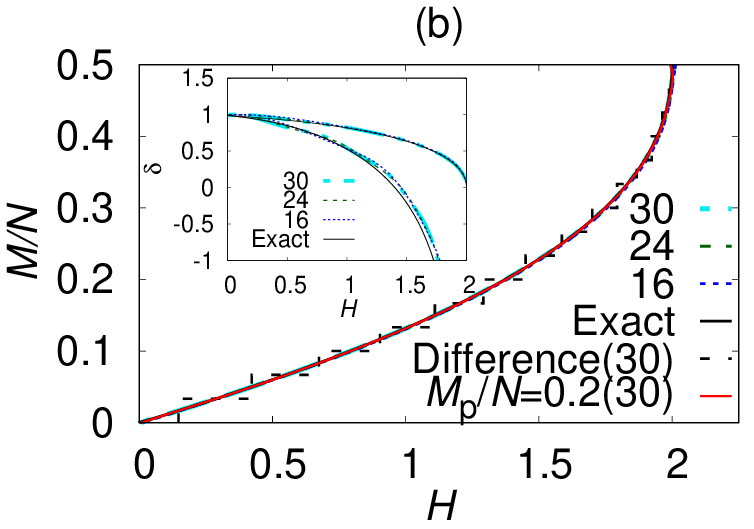}}
  \resizebox{0.40\textwidth}{!}{\includegraphics{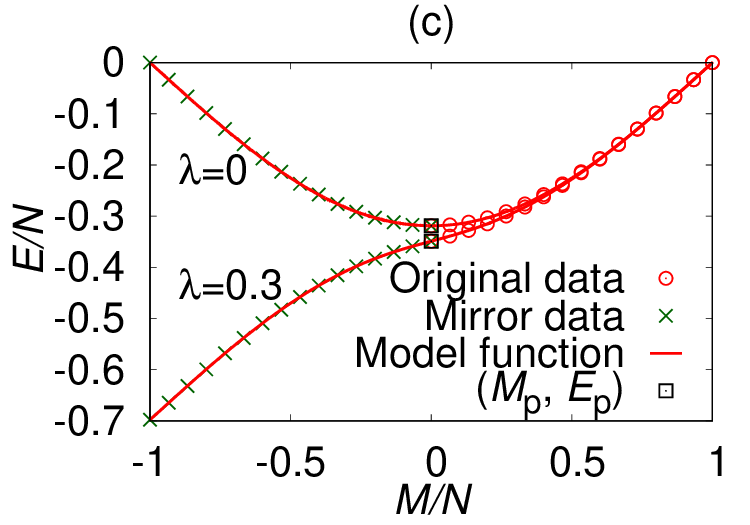}}
  \caption{
B-GK results for 
the magnetization curve. 
Results of 
the difference and the exact solution 
are plotted for comparison.
The inset shows model functions of the exponent $\delta$ 
obtained by both expressions in equation~(\ref{eq:delta}).
We discarded data of $|M-M_{\rm p}|$ smaller than twice the standard
deviation of $M_{\rm p}$.
(a)
The dimerized AF XY chain with $N=30$.
(b) 
The AF Heisenberg spin chain with $N=30, 24$, and $16$.
A red line shows
the magnetization curve when a value of $M_{\rm p}/N$ is fixed 
to 0.2 in the B-GK analysis.
(c) The ground-state energy $E(M)$ of the dimerized AF XY chain with 
$\lambda=0$ and $0.3$. 
Mirror data for $\lambda=0$ is symmetric($A_{\rm s}=1.0000(1)$),
whereas those for $\lambda=0.3$ is antisymmetric($A_{\rm s}=-0.9926(6)$).
}
\label{fig:mh}
\end{figure}

Our B-GK results for the magnetization curve 
and the ground-state energy
are plotted in Fig.~\ref{fig:mh}.
The magnetization curve obtained by our method coincides with the
exact result accurately in the whole range of the magnetic field.
We only notice a small disagreement in the vicinity of $M=0$ 
when $\lambda$ is very small in the XY model.
The agreement is particularly excellent when the spin gap is large or zero.
We did not observe finite-size effects in the AF Heisenberg chain.
The average deviation of the magnetization curve from the exact one was
0.0007(8) for $N=30$ and 0.003(2) for $N=16$ in the range of $0<H<1.99$,
where the error is estimated by the standard deviation among two thousand
data points.

The model function for the critical exponent $\delta$ is plotted 
in insets of Fig.~\ref{fig:mh}.
Two model functions with equivalent expressions
converge to the same value.
The exact results in the gapful models 
(black solid lines of $\lambda=0.1, 0.3, 0.5$ in Fig.~\ref{fig:mh}(a))
approach 2 with a finite slope,
whereas
those of the gapless models 
(a black solid line of $\lambda=0.0$ in Fig.~\ref{fig:mh}(a) and 
that in Fig.~\ref{fig:mh}(b))
approach 1 with a zero slope.
This difference in the exponent also clearly identified
the gapful and the gapless ground states. 
Our B-GK result exhibits an agreement with the exact one except for the
case of the $\lambda = 0.1$ XY model.
It deviated from the exact curve before reaching the critical field and
exhibited a non-uniform behavior.
We consider it as an outcome of the small lattice size compared to 
the magnitude of the spin gap.

We here present an alternative estimate for the spin gap 
by the B-GK method.
Namely,
the critical magnetic field at the zero magnetization,
$H(0)=\partial E(M)/\partial M|_{M=0}$, 
corresponds to the spin gap above the ground state.
The conventional estimate for the spin gap, $E(1)-E(0)$, is regarded
as an approximation of the numerical differentiation by the difference.
Since $H(0)$ is the slope of $E(M)$ at $M=0$, 
the symmetric mirror ($A_{\rm s}=1$) of the data are 
favored when the spin gap is zero, whereas the
asymmetric mirror ($A_{\rm s}=-1$) of the data are 
favored when it is finite; see  Fig.~\ref{fig:mh}(c).
Therefore,
we expect to judge the ground state 
whether gapless or gapful by the estimate of this parameter.

The estimates of the spin gap and the asymmetric parameters 
are summarized in Tab.~\ref{tab:param}.
This algorithm easily identified the large-gap state and the gapless state.
In the gapless models (the XY model with $\lambda=0$
and the AF Heisenberg model),
the estimate of the spin gap according to the critical field $H(0)$
almost vanishes even in a finite lattice with 30 spins.
The asymmetric parameter $A_{\rm s}$ 
took a value close to $1$ in the gapless state and 
a value close to $-1$ in the gapful state 
(the XY model with $\lambda \ge 0.1$).
This parameter indeed identified by its score
whether the ground state is gapless or gapful in the thermodynamic limit.
The estimate took a value away from $\pm 1$ for $\lambda = 0.1$ and $0.05$.
These ambiguous results 
together with the non-uniform behavior of the critical exponent 
suggest that the algorithm needs more data 
with larger lattice sizes to clearly identify the state as gapful.

In the analyses of the spin gap estimation, we fixed the value of $M_{\rm p}$
to 0 as prior information for the Bayesian inference,
because we know that the ground state of these systems lie in the $M=0$ sector.
Here, a question arises.
What happens if we feed an incorrect prior information 
to the Bayesian inference?
Does the present method give a continuous magnetization curve
if we set $M_{\rm p}$ at a value where there is no magnetic plateau?
We here show that the algorithm produces only a negligibly narrow plateau
at a wrong value of magnetization.
We checked the method by setting $M_{\rm p}/N=0.2$ 
in the AF Heisenberg spin chain. 
We obtained
the magnetization curves both below and above this $M_{\rm p}$ value
individually
by the same procedure as in the
incommensurate modulated model in Fig.~\ref{fig:income}.
The results are plotted in Fig.~\ref{fig:mh}(b) by a red line. 
We estimated
two critical field above and below the magnetization plateau at
1.37862(8) and 1.37762(9), respectively; 
in other words, the plateau width was only 0.001.
Asymmetric parameters were 0.883(2) and 1.143(2), which are close to the
gapless value 1.
We did not identify the difference of the magnetization curve from the exact
result. 
The average deviation of the magnetization curve from the exact one was 
0.0006(7) in the range of $0<H<1.99$, which is consistent with the results 
obtained by fixing $M_{\rm p}=0$.
The error is estimated by the standard deviation among four thousand
data points.
In short,
the algorithm answers with practically no plateau
if we give a value of $M_{\rm p}$
at which there is no magnetization plateau.

\subsubsection{Size extrapolation of the spin gap}

An estimate for the spin gap $\Delta$ by $H(0)$
always gives a value smaller 
than the conventional estimate $E(1)-E(0)$ 
because $E(M)$ is a convex function.
Both estimates should converge to the same value in the
thermodynamic limit accompanied by different size dependences.
A combined size-extrapolation analysis
using both $E(1)-E(0)$ and $H(0)$
gives an improved estimate for the spin gap in the thermodynamic limit.

We can perform this extrapolation analysis by the
same procedure as we fixed the plateau point using the mirror data.
Namely,
we plot spin-gap data against $1/N$ and 
search for an extrapolated point $(1/N=0,\Delta)$ by the Bayesian inference
such that a model function connecting the
original data and the mirror data becomes the smoothest.
We now have two estimates for the spin gap, $H(0)$ and $E(1)-E(0)$.
For each data series, we define the mirror data with respect to the 
common extrapolated point $(1/N=0, \Delta)$.
We estimate the log-likelihood function of the model function for each 
data series, and just sum them up.
We search for a value of $\Delta$ that yields the largest value of the sum.
Once we obtain the spin gap and the hyperparameters for each data series,
we have a model function by equation~(\ref{eq:modelfunc}).

The results are summarized in Tab.~\ref{tab:param} compared with those
obtained by the quadratic least-squares method.
Namely, $\Delta(N)=\Delta + b/N^2 + c/N^4$ in the gapful models, and
$\Delta(N)=\Delta + b/N + c/N^2$ in the gapless models and in the kagome
antiferromagnet.
The B-GK method gave a result consistent with the least-squares one
regardless of whether the system is gapless or gapful.
The extrapolated values listed in Tab.~\ref{tab:param} remained the same 
even if we changed the horizontal axis
of the plot from $1/N$ to $1/N^2$ except for the case of the
kagome antiferromagnet. 
We changed the horizontal axis to 
$1/\sqrt{N}$ and the result changed from 0.028(1) to 0.080(2).
The extrapolation was sensitive
because we only had three data for the size extrapolation.

\begin{table*}
\hfil
\begin{tabular}{|l|l|l|l|l|l|l|}
\hline
Model & $A_{\rm s}$ & \multicolumn{2}{c|}{Spin gap ($N=30$)} & \multicolumn{3}{c|}{Spin gap ($N=\infty$)} \\
 &  & $H(0)$& $E(1)-E(0)$ & B-GK & Least squares & Exact\\
\hline
XY: $\lambda=0.50$ & $-0.9994(1)$ & $0.49720(3)$&0.5000000 &0.49999(2) &0.50009(2) & 0.5\\
%\hline
XY: $\lambda=0.30$ & $-0.9926(6)$ & $0.29397(4)$&0.3000156 &0.29978(5) &0.30016(4)  & 0.3 \\
%\hline
XY: $\lambda=0.10$ & $-0.476(6)$  & $0.0778(1)$ &0.1050752 &0.0998(9)  &0.0939(2) & 0.1 \\
%\hline
XY: $\lambda=0.05$ & $0.350(8)$   & $0.0251(2)$ &0.0678188 &0.04564(2) &0.0499(5) & 0.05\\
%\hline
XY: $\lambda=0.00$ & $1.0000(1)$  & $0.00001(7)$&0.0524066 &0.000000(4)&-0.0005(9) & 0\\
\hline
1D-AFH & $0.946(2)$ & $0.0051(2)$  &0.1471492 & 0.00179(5) &0.0020(1) & 0\\
\hline
kagome-AFH& $0.69(3)$ & $0.048(6)$  &0.1509240 & 0.028(1) &-0.158 & \\
\hline
\end{tabular}
\caption{
Our estimates of the
asymmetric parameter and the spin gap.
The magnitude of the spin gap according to
the critical field $H(0)$, and that according to the 
conventional definition $E(1)-E(0)$
for each model when $N=30$
and the estimates after the size extrapolation ($N=\infty$)
by the B-GK method using both $H(0)$ and $E(1)-E(0)$
and that by the quadratic least-squares method using only $E(1)-E(0)$.
Errors of the B-GK results are estimated as the standard deviation among
40 best results of the log-likelihood function out of 800 
random-search trials.
}
\label{tab:param}
\end{table*}

\begin{figure}
\hfil
  \resizebox{0.45\textwidth}{!}{\includegraphics{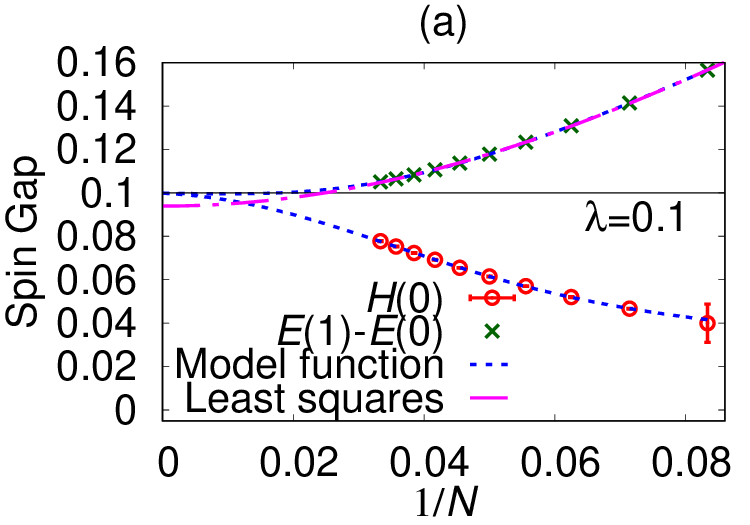}}
  \resizebox{0.45\textwidth}{!}{\includegraphics{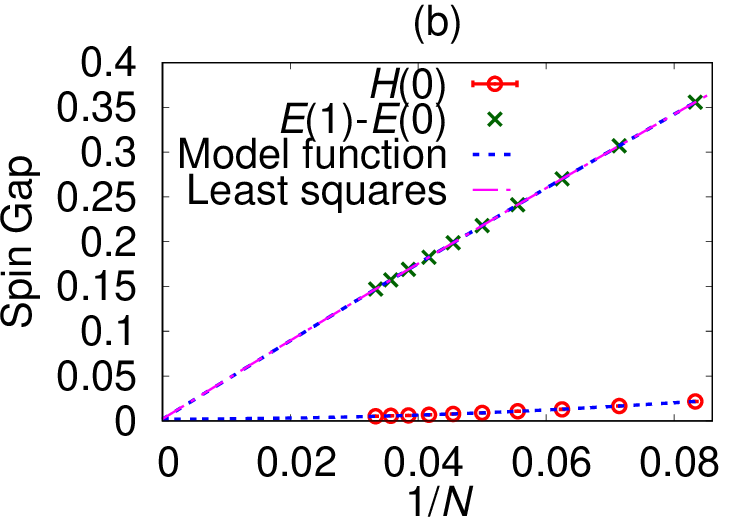}}
  \caption{
B-GK results for the
size extrapolation of spin-gap data.
The result of the quadratic
least-squares method is plotted for comparison.
(a) 
The dimerized AF XY chain when $\lambda=0.1$.
(b)
The AF Heisenberg spin chain.
}
\label{fig:gap}
\end{figure}

The model function for the size extrapolation is plotted in Fig.~\ref{fig:gap}.
It converged to a finite value with a zero slope in a gapful model.
It is consistent with a reasoning that there is a typical length scale $\xi$,
beyond which the data loses the size dependence.
It corresponds to a spin gap $\Delta$ as $\xi \sim 1/\Delta$.
The model function in the AF Heisenberg spin chain converged 
to zero with a finite slope, which is also consistent with this reasoning.

\subsection{Application to the kagome antiferromagnet.}

Results presented above guaranteed that the present method gives the
magnetization curve and the spin gap accurately.
The asymmetric parameter $A_{\rm s}$ 
and the model function for the exponent $\delta$ are
expected to identify
whether the spin gap is zero or finite.
Now, we apply this method to the $S=1/2$ AF Heisenberg model on the
kagome lattice.
We first focus on the magnetization curve of this model, particularly
on the existence of the $1/9$-plateau.
It was claimed to exist by Nishimoto {\it et al.}~\cite{nishimotokagome},
whereas denied by Nakano and Sakai~\cite{nakano18}.

\begin{figure*}
\hfil
  \resizebox{0.45\textwidth}{!}{\includegraphics{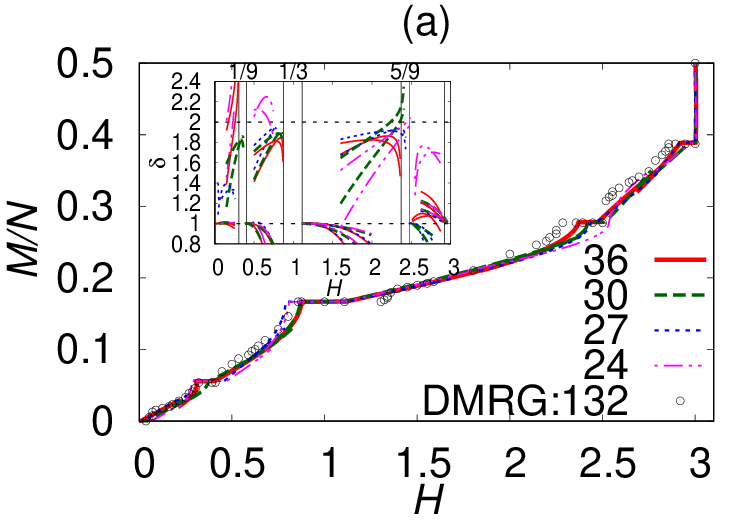}}
  \resizebox{0.45\textwidth}{!}{\includegraphics{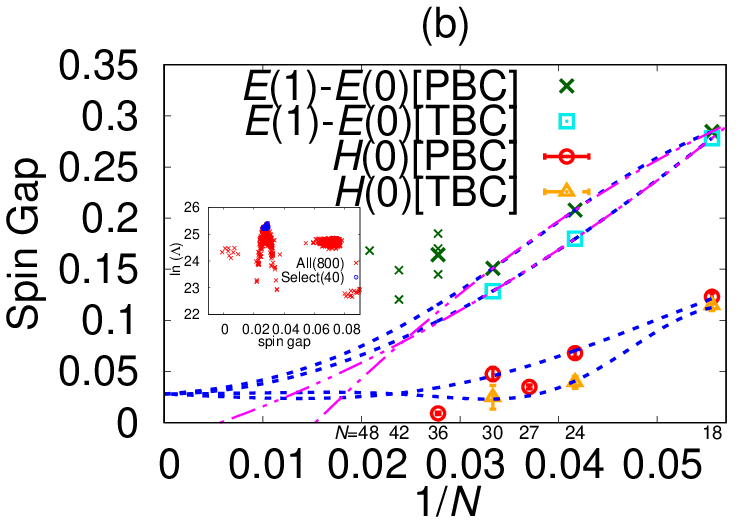}}
  \resizebox{0.60\textwidth}{!}{\includegraphics{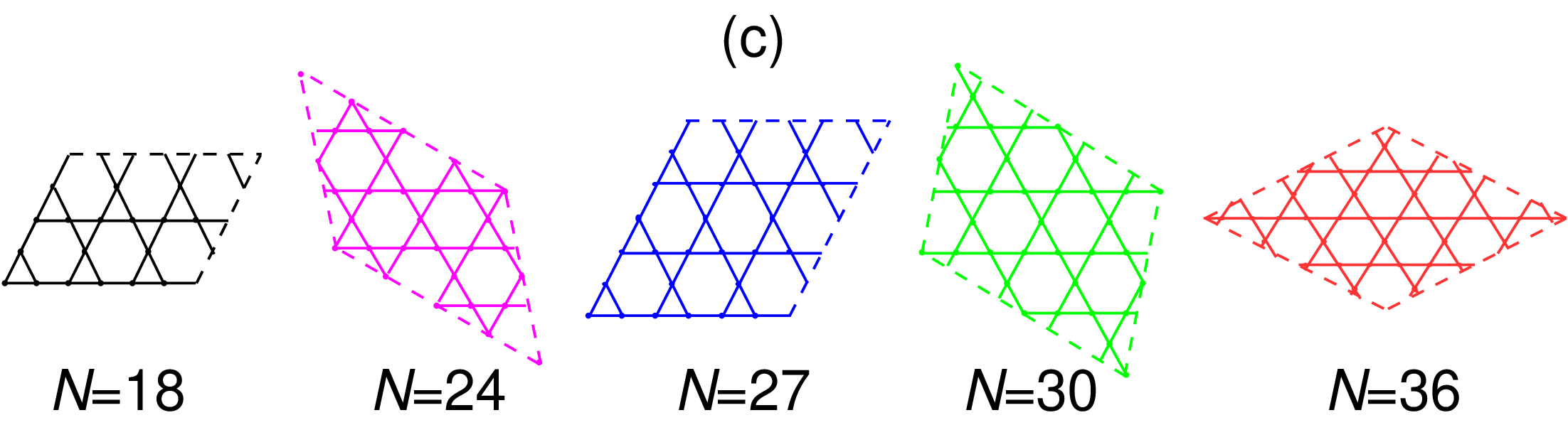}}
\caption{
B-GK results of the kagome antiferromagnet.
(a) Magnetization curves obtained from the ED data
with $N=$ 36, 30, 27, and 24. 
The DMRG result is taken from Ref.~\cite{nishimotokagome}, in which the
spin number is $132$.
The inset shows model functions of the exponent $\delta$
obtained by both expressions in equation~(\ref{eq:delta}).
We discarded data of $|M-M_{\rm p}|$ smaller than twice the standard
deviation of $M_{\rm p}$.
(b) Our size extrapolation of spin-gap data.
Data denoted by PBC (TBC) are results using the 
periodic (twisted) boundary conditions.
Model functions of the B-GK analysis (blue) as well as the
quadratic least-squares fitting results (magenta) are plotted
with lines.
We also plotted with smaller symbols
data of $E(1)-E(0)$ for larger and different-shape lattices
taken from Refs.\cite{lauchli} and \cite{lauchli2}.
The inset shows a distribution of the log-likelihood function for
each extrapolated spin gap.
We took
the average of 40 best results among 800 trials 
and put an error bar by the standard deviation.
The result was $0.028(1)$.
(c) Shapes of kagome lattice treated in this analysis.
}
\label{fig:kagome}
\end{figure*}

We calculated the ground-state energy for $N=$
36, 30, 27, and 24, and applied the 
B-GK analysis to the data.
The ground-state energy for $N=36$ were obtained by the numerical package 
${\cal H}\Phi$ \cite{misawa1}.
The results for the magnetization curve are 
shown in Fig.~\ref{fig:kagome}(a).
Here, we used five values of 
$2 M_{\rm p}/N$ at $0$, $1/9$, $3/9$, $5/9$ and $7/9$
as prior information and performed
each search for $E_{\rm p}$ without using the data point of 
$(M_{\rm p},E(M_{\rm p}))$.
We could not estimate the plateau point without fixing the value 
as was done in the modulated spin chain in Fig.~\ref{fig:income}.
This is just because we only have a few data points between the neighboring
magnetic plateaux for these lattice sizes.
The magnetization curve for each plateau connects with each other smoothly.
They are consistent with
the one obtained by the DMRG calculation~\cite{nishimotokagome} with 132 spins,
although the
locations of some of the plateau edges differ presumably owing
to finite-size effects.
We observed the 1/9-plateau clearly except for the size $N=30$.
The plateau width exhibits strong finite-size effects.
The situation is similar for other magnetic plateaux.
The 5/9-plateau vanished only for the size $N=24$.
The plateau magnetic states with the broken translational symmetry
may not fit properly in such distorted lattices.
Shapes of the finite-size kagome lattice are summarized in Fig.~\ref{fig:kagome}(c).
A result for a symmetric lattice mostly exhibits wide magnetic plateaux.

The model function of the critical exponent $\delta$ for each system size
is shown in an inset of Fig.~\ref{fig:kagome}(a).
It approaches a value
around 2
in the low-field side of the plateaux of $2M_{\rm p}/N=1/9$, $3/9$, and $5/9$.
However, most of the data exhibit a crossover bending
before reaching the critical point. We observed the similar behavior in
an inset of Fig.~\ref{fig:mh}(a),
where the lattice size was too small for the spin gap
in the XY model with $\lambda=0.1$.
The exponent in the high-field side of the $1/9$- and the $1/3$-plateaux 
as well as that of the ground state at $M_{\rm p}=0$ 
converges to one with a zero slope, which is similar to the gapless
models in one dimension.
The result in the high-field side of the $5/9$-plateau exhibited 
the strong finite-size effect,
which suggests that we need more data with larger lattice sizes.
The exponent in the low-field side of the $7/9$-plateau converges to one
but from above with a finite slope, which is a behavior different from 
the other plateau edges.
These differences in the behavior of $\delta$ suggest different
scenario for the excited states at each plateau edge.
The estimated critical exponent in the low-field side of the $1/3$-plateau is
 consistent with the estimate 
obtained by Sakai and Nakano~\cite{sakai11},
which gave $\delta=1.92(99)$,
but 
that in the high-field side
deviated from their estimate, which gave $\delta=0.56(15)$.

The size extrapolation of the spin gap for the kagome antiferromagnet 
is shown in Fig.~\ref{fig:kagome}(b).
We also plotted for comparison
the spin-gap data for larger and different-shape 
lattices\cite{lauchli,lauchli2}.
Here, we obtained the spin-gap data for systems with 18, 24, and 30 spins
under the periodic boundary condition as well as under
the twisted boundary condition with a $\pi$ flux.
By using four data series with a common extrapolated point,
we estimated the gap at $0.028(1)$.
The spin-gap data for $N=36$ (both $E(1)-E(0)$ and $H(0)$)
and $N=27$ (only $H(0)$)
as well as those for larger and different-shape lattices
much deviate from the trend of other
system sizes, and hence we did not use them in the extrapolation analysis.
The extrapolated gap shifts to 0.078(3) if we include the 
lowest gap data of $N=36$ and 42 among the shape-dependent results.
In this analysis, we needed to set the error bars of the ED data to 
0.003 in order to accept the deviated data to be fitted smoothly.
The extrapolation analysis fails if we include the data of $N=48$ or
use other data of $N=36$ and 42.

All data (both $E(1)-E(0)$ and $H(0)$) up to the present lattice sizes
seem to decrease faster than the behavior of linear convergence to zero.
Since the spin gap cannot be negative, there must exist a size-crossover
point at which the trend changes.
This is also consistent with the behavior of 
each model function in Fig.~\ref{fig:kagome}(b), which almost
lost the size dependence before reaching the thermodynamic limit.
These results suggest that the gap is finite,
although
inclusion of large-size data may affect the final result much, 
which is always the case with the extrapolation analysis.
Considering the size dependence and the shape dependence of the spin-gap data
obtained up to the present, we need data for much larger lattices.

Our extrapolated value of the spin gap
is consistent with three previous results:
the DMRG calculation on a kagome strip~\cite{heprx}, 
which gave 0.02-0.04;
the DMRG calculation using the lattice deformation~\cite{nishimotokagome}, 
which gave 0.05(2);
the experimental estimate~\cite{expkagome}, 
which gave 0.03-0.07.
Meanwhile, it physically contradicts
recent results~\cite{liao,heprx} that suggest a gapless Dirac spin liquid,
although our result of the model function of the exponent $\delta$ 
converging to 1 supports this picture.
The existence of the spin gap thus remains under debate.

\section{Discussion}

We can obtain a continuous and differentiable model function 
of data by the Gaussian-kernel method.
The Bayesian inference also gives an exclusion point of data very accurately.
These two points are the essential ingredient 
for the success of the present method.
The accuracy of each physical quantity
was much higher than the resolution due to the finite lattice size.
It is a great advantage when we only have insufficient data.
Applications to incommensurate systems 
and to two-dimensional quantum systems are promising.
A critical exponent can be obtained 
as a continuous curve converging to its critical value at the
critical point.
It is an alternative method for estimating critical exponents
to the conventional scaling analysis.

Although the original Gaussian-kernel regression is 
basically the data {\it interpolation},
we extended it to the data {\it extrapolation} by defining mirror data. 
There remains a room for improvement in defining the mirror data 
and for another attempt to make this extension, depending on the problem.
In the present data extrapolation,
we do not need to assume any form of a model function as has
been done in the least-squares method.
The method automatically finds the most probable model function 
converging to the extrapolated point.
We can avoid ambiguity due to the choice of the function form.
We checked in this paper 
that the B-GK results are robust against the change of the horizontal axis.
The present method may replace a conventional data extrapolation and 
numerical differentiation,
which have been applied to various scientific analyses.

The asymmetric parameter introduced in the mirror data
and the model function of the exponent $\delta$
possibly serve 
as an identifier for a gapless/gapful system.
The former one identifies it by
its values $A_{\rm s}\simeq 1$ 
or $-1$.
The latter one does it by its converging value and behavior.
Systems investigated in this paper satisfied these criteria.
In a case with a very small spin gap, 
the algorithm suggested that it is not conclusive by giving 
an ambiguous value and the behavior.
It indicates the necessity of preparing more data with larger lattices.
In other words, 
the algorithm did not make a mistake to identify
the small-gap system as gapless.
We may regard this as a partly success.

In this context,
the spin-gap issue in the kagome antiferromagnet remained unsettled.
The size extrapolation of spin-gap data gave a very small gap,
whereas
a model function of the exponent $\delta$ in an inset of 
Fig.~\ref{fig:kagome}(a) showed a convergence to 1,
which is common to the gapless systems.
Finally, 
the asymmetric parameter took an ambiguous value between $1$ and $-1$.
This contradiction reflects the difficulty in the study of the 
kagome antiferromagnet.
The lattice sizes treated in this paper were restricted to be small.
It may be much smaller than a typical size for which 
physics of the kagome antiferromagnet clearly emerges.
We need careful analyses beyond the lattice sizes that we treated,
because the size effect in the kagome antiferromagnet sometimes changes its
trend depending on the shape of lattice~\cite{nakano11b,lauchli}.

Let us make one comment.
The present algorithm is not universally better than other algorithms
for any problems.
This is because of the no-free-lunch theorem~\cite{rinko,goodfellow} 
in machine learning.
There may exist another algorithm that works better for some problems.
We found that the B-GK method works better than 
numerical differentiation by the difference and better than
the least-squares extrapolation.
On the other hand,
we needed to give some prior information or directions
for the algorithm to work well.
We utilize this information 
depending on each problem and data~\cite{goodfellow}.
It is a general aspect of the machine-learning algorithm.
Therefore, we cannot use it as a black box.
A disadvantage of the kernel method compared to the neural-network
algorithm is the computational cost $O(d^2)$, 
where $d$ is the number of data.
However, the computational ability is still increasing and it
may solve the disadvantage.

\section{Method}

In order to obtain the magnetization curve in the ground state numerically, 
we first evaluate the ground-state energy $E(M)$
in each total magnetization subspace with $M=\sum_i S_i^z$.
The relation between the magnetization and the magnetic field, $H$, is
given by
\begin{equation}
H(M)=\frac{\partial E(M)}{\partial M}.
\label{eq:HM}
\end{equation}
We plot $M$ against $H(M)$ to obtain the magnetization curve.
This differentiation has been estimated by the difference as in
$H(M)=E(M)-E(M-1)$,
which exhibits a stepwise behavior.
We cannot determine the existence of a nontrivial magnetic plateau
from such results.
The DMRG method is a good choice
to increase the system size in low-dimensional systems,
but
the differentiation still has been performed by the difference.
The resolution of the result is limited by the lattice size.
This situation is common in numerical differentiation 
in various fields up to the present.
The B-GK method replaces it
by analytic differentiation of the model function.

Let us briefly explain the 
Gaussian kernel regression~\cite{bishop,rinko,goodfellow,harada}.
We try to obtain a model function for data 
$(x_i, y_i, \delta y_i)$, where $i=1, 2, \cdots ,d$ is the data index,
and $\delta y_i$ denotes the error of datum $y_i$, e.g. a Monte Carlo error.
The point is to use
the Gaussian kernel function $K(x_i, x_j)$ for $x_i\ne x_j$ of the form
\begin{equation}
K(x_i, x_j)=\theta_1^2 
\exp\left[-\frac{(x_i-x_j)^2}{2\theta_2^2}\right]+\theta_3^2,
\label{eq:kernel}
\end  {equation}
where $\theta_1$, $\theta_2$ and $\theta_3$ are hyperparameters,
which are fixed by maximizing the log-likelihood function
\begin{equation}
 \ln (\Lambda)=-\ln|C|- \sum_{i,j}^d y_iC^{-1}_{ij} y_j,
\label{eq:logli}
\end  {equation}
where 
$C$ is a $d$-dimensional covariance matrix defined by
\begin{equation}
C_{ij}=[\delta y_i]^2\delta_{ij}+K(x_i, x_j),
\label{eq:Cij}
\end  {equation}
and
$|C|$ denotes the determinant of $C$.
This maximization problem is solved by a numerical 
package~\cite{numericalrecipe}, such as
the conjugate-gradient method or the downhill simplex method.
With the estimates of the  hyperparameters,
we have a continuous and generally nonlinear 
model function for arbitrary $x$ as
\begin{equation}
{\cal F}(x)=\sum_{i,j}^d K(x,x_i)(C^{-1})_{ij} y_j.
\label{eq:modelfunc}
\end  {equation}
This function consists of data $\{ y_j\}$ with their weights
given by the Gaussian kernel function.
We may roughly consider that
the weight is larger if the data is closer to $x$ or the error bar is smaller.
Since $x$ appears only in $K(x, x_i)$, we can find the derivatives 
of ${\cal F}(x)$ by replacing $K(x, x_i)$ with $\partial K(x, x_i)/\partial x$
and its higher derivatives.

Now, suppose that we have a set of energy data, $(M_i, E_i)$, as shown in
Fig.~\ref{fig:income}(a).
When the magnetic plateau exists in the magnetization curve, 
the function
$H(M)$ of equation~(\ref{eq:HM}) should jump at $M=M_{\rm p}$
and the data $(M_i, E_i)$ should exhibit a sudden change of slope 
at the plateau point.
Because of this, we cannot model all the energy data 
by one smooth function that straddles the plateau point.
This is the same situation in which we determined the phase
transition temperature in the two-dimensional Ising model
by the B-GK method~\cite{totabayes}.
Energy data were not modeled by a smooth
function beyond the phase transition temperature.
The critical temperature was estimated as an exclusion point.
Here, we follow the same recipe.

We first split the energy data
into the high-field region ($M>M_{\rm p}$)
and the low-field region ($M<M_{\rm p}$)
at a plateau point $(M_{\rm p}, E_{\rm p})$,
which is for the moment unknown and is
to be estimated by the Bayesian inference.
We start the inference from a random initial value of
$(M_{\rm p}, E_{\rm p})$.
We set $d_{\rm h}$ pieces of points $(x_i, y_i)$ from the
data in the high-field region as
\begin{equation}
x_i=(M_i-M_{\rm p})/N,~~~ y_i=(E_i-E_{\rm p})/N
\label{eq:orig}
\end{equation}
for $i=1, 2, \cdots, d_{\rm h}$,
where $d_{\rm h}$ is the number of data points in the high-field region.

A key of the present algorithm is to introduce additional data points 
$(x_{i+d_{\rm h}}, y_{i+d_{\rm h}})$ as 
the mirror data 
as shown in Fig.~\ref{fig:income}(a) by cross symbols.
Here, we consider a line $y=ax$ that goes through the plateau point with a
slope $a=(y_{\rm u}-y_{\rm \ell})/(x_{\rm u}-x_{\rm \ell})$, where
$(x_{\rm u}, y_{\rm u})$ and $(x_{\rm \ell}, y_{\rm \ell})$ 
are the upper and the
lower neighboring data of the plateau point, and hence it gives
an approximate slope at the plateau point.
We consider mirror data with respect to this line.
Namely,
\begin{eqnarray}
x_{i+d_{\rm h}}&=&-x_i, \nonumber \\
y_{i+d_{\rm h}}&=& A_{\rm s}(y_i-ax_i)+ ax_{i+d_{\rm h}} \nonumber \\
&=&A_{\rm s} y_i -(A_{\rm s}+1)ax_i
\label{eq:mirror}
\end{eqnarray}
for $i=1, 2, \cdots, d_{\rm h}$, where $A_{\rm s}$ is an additional 
hyperparameter to be estimated by the Bayesian inference.
The mirror is antisymmetric if $A_{\rm s}=-1$ and symmetric if $A_{\rm s}=1$.
This is an adoption of the method of mirror images in electrostatics;
mirror charges are fixed so that the electric field may continue smoothly
at the boundary.
The parameter $A_{\rm s}$ is determined so that the model function may become
as smooth as possible, which is automatically realized by 
the Gaussian kernel regression.
We searched for $A_{\rm s}$ from the initial value randomly distributed 
between $-1$ and $1$.

We introduce the asymmetric parameter $A_{\rm s}$
because of the following reason.
In general data regression,
the quality of modeling becomes poorer near the edges of data series.
Since the magnetic plateau exists between two edges of the $E(M)$ data series,
the numerical precision of the obtained model function can deteriorate much.
Then, its derivative, the magnetization curve, would exhibit 
numerical instability near the plateau point.
The introduction of the mirror data solves this problem by making the
data edge the midst point.

Using $2d_{\rm h}$ data points in equations~(\ref{eq:orig}) 
and (\ref{eq:mirror}),
we evaluate the log-likelihood function~(\ref{eq:logli})
in the high-field region.
The log-likelihood function
in the low-field region is also evaluated by applying the same procedure
to $2d_{\ell}$ data points, 
where $d_{\ell}$ is the number of data points in the low-field region.
We searched for a common estimate of $(M_{\rm p}, E_{\rm p})$
and individual estimates of $A_{\rm s}$ 
in both low- and high-field regions
so that the sum of the two log-likelihood functions may take the maximum.
We carried out this search for 800 times.
Then, we took the average over 40 best results
and put an error by the standard deviation.
When the distribution of the log-likelihood function is broad,
we increased the search up to 1600 times and
took the average over 400 best results.
This occurred when we estimated the magnetization
curve of the kagome antiferromagnet.
Using the estimated parameters,
we obtain a model function of $E(M)$ 
in the form of equation~(\ref{eq:modelfunc})
for each region.
We readily have a model function of $H(M)$ by the derivative of $E(M)$.  

The critical exponent of the magnetization at the plateau edge $H_{\rm p}$
is defined by 
$(H-H_{\rm p})\sim (M-M_{\rm p})^{\delta}$.
By logarithmic differentiation and the l'H\^opital's rule,
we can estimate this exponent in the following expressions:
\begin{eqnarray}
\delta&=&
\lim_{M\to M_{\rm p},
      H\to H_{\rm p}}
\frac{M-M_{\rm p}}{H-H_{\rm p}}
\frac{\partial H}{\partial M} \nonumber \\
&=&
\lim_{M\to M_{\rm p}}
1+
\frac{M-M_{\rm p}}{\frac{\partial H}{\partial M}}
\frac{\partial^2 H}{\partial M^2}
\label{eq:delta}
\end  {eqnarray}
We have 
model functions
 of 
these expressions by analytic differentiation of
the model function of $H(M)$.
By plotting two model functions with equivalent expressions,
we can fix the common extrapolated point as the critical exponent.
The second expression 
trivially takes a value 1 if we set $M=M_{\rm p}$ as long as the 
denominator $\partial H/\partial M$ is not exactly zero.
We observe it as a crossover bending of the model function. 
Before reaching the critical point,
both expressions exhibit numerical instability 
in the vicinity of $M=M_{\rm p}$.
It is caused by the situation 0/0 for the first expression, and is caused
by the poor estimation of the derivatives for the second expression; 
see an inset of Fig.~\ref{fig:income}(b).
We need to discard these data 
where the value of $|M-M_{\rm p}|$ is smaller than several times the standard 
deviation of $M_{\rm p}$.

When we apply this method for the size extrapolation of the spin gap,
we set $x_i=1/N_i$ and $y_i=\Delta(N_i)$, where $\Delta(N_i)$ is the
spin gap estimated at the size $N_i$.
We search for the extrapolated point $(1/N=0, \Delta)$ by defining the
mirror data in the negative side of $1/N$.
We set the approximate slope $a=0$ in this analysis because there is no
real data in the lower side of the extrapolated point.

It is very important to avoid overfitting and underfitting
in the machine-learning algorithm~\cite{bishop,rinko,goodfellow}. 
When overfitting occurs,
the algorithm tries to fit data strictly, ignoring a trend of whole data.
It occurred mostly in the analysis of the kagome antiferromagnet.
For example,
an estimated spin gap $H(0)$ took a negative value 
even for a finite lattice size.
In another case,
the extrapolated spin gap took a value larger than the finite-size data.
The cross validation is generally useful to avoid the overfitting.
In the textbook example, we would randomly select
$(d-1)$ pieces of data out of $d$ and search for the parameters.
The estimated parameters would be validated 
with another choice of $(d-1)$ pieces of data.
This cross validation actually fails when $d$ is very small.
In the present analysis, we introduced another method of
cross validation by using random noise.
When we estimated parameters, we first added random noise to
the original data, where the noise was set proportional to the error
of the data.
We validated
the estimated parameters by calculating the log-likelihood
function using data with different random noise.
We found that this procedure works fine even when the number of data is 
very much restricted.
We mostly set noise amplitude to $10^{-6}$
since the ED data are supposed to be numerically exact.
We needed to increase it up to an order of $10^{-4}$ 
when strong overfitting occurred.
On the other hand,
underfitting may occur giving just a flat line of averaged data,
if the noise was set too large.
We needed to check the results and set values of the noise of the 
original data so that both overfitting and underfitting may vanish.
This is the only procedure that we did manually.

 \section{Acknowledgements}
 The author would like to thank Dr.~Kiyomi Okamoto, Dr.~Chisa Hotta, 
 Dr.~Katsuhiro Morita, and Dr.~Naomichi Hatano for valuable discussions
 and comments.
 He also thanks Dr.~Takahiro Misawa for providing him with the numerical data.
 The use of the ED package TITPACK ver.2 programmed
 by Dr.~Hidetoshi Nishimori is gratefully acknowledged.
 This work was partly supported by the joint research program of
 Molecular Photoscience Research Center, Kobe University.

 \end{document}